\documentclass[pra,twocolumn,superscriptaddress,amsmath,amsthm,showpacs]{revtex4}
\usepackage{amssymb}
\usepackage{graphicx}

\newcommand{\bra}[1]{\langle #1 |}
\newcommand{\ket}[1]{| #1 \rangle}
\newcommand{\braket}[2]{\langle #1 | #2 \rangle}
\newcommand{\ketbra}[2]{| #1 \rangle \langle #2 |}
\newcommand{\expect}[1]{\langle #1 \rangle}

\begin{document}
\title{Demonstrating nonclassicality and non-Gaussianity of single-mode fields: Bell-type tests using generalized phase-space distributions}
\author{Jiyong Park}
\affiliation{Department of Physics, Texas A\&M University at Qatar, Education City, P.O.Box 23874, Doha, Qatar}
\author{Hyunchul Nha}
\affiliation{Department of Physics, Texas A\&M University at Qatar, Education City, P.O.Box 23874, Doha, Qatar}
\affiliation{School of Computational Sciences, Korea Institute for Advanced Study, Seoul 130-722, Korea}
\date{\today}

\begin{abstract}
We present Bell-type tests of nonclassicality and non-Gaussianity for single-mode fields employing a generalized quasiprobability function. 
Our nonclassicality tests are based on the observation that two orthogonal quadratures in phase space (position and momentum) behave as independent realistic variables for a coherent state. Taking four (three) points at the vertices of a rectangle (right triangle) in phase space, our tests detect every pure nonclassical Gaussian state and a range of mixed Gaussian states. These tests also set an upper bound for all Gaussian states and their mixtures, which thereby provide criteria for genuine quantum non-Gaussianity. We optimize the non-Gaussianity tests by employing a squeezing transformation in phase space that converts a rectangle (right triangle) to a parallelogram (triangle), which enlarges the set of non-Gaussian states detectable in our formulation. We address fundamental and practical limits of our generalized phase-space tests by looking into their relation with decoherence under a lossy Gaussian channel and their robustness against finite data and non-optimal choice of phase-space points. Furthermore, we demonstrate that our parallelogram test can identify useful resources for nonlocality testing in phase space.
\end{abstract}

\maketitle

\section{Introduction}
Describing a quantum state in phase space \cite{Wigner1932} is a very useful approach to make a comparison between quantum mechanics and classical mechanics. It provides a valuable insight into the phenomenon of quantum-to-classical transition \cite{Zurek2003} and a powerful tool to manifest nonclassical effects in quantum optics \cite{Barnett} and continuous variable (CV) quantum informatics \cite{Braunstein2005, Weedbrook2012}. One of the remarkable distinctions between quantum and classical phase-space distributions is that a negative value is allowed for a quantum state. Although the negativity in phase space thus demonstrates nonclassicality immediately, it enables us to detect only a limited subset of nonclassical states. There exist nonclassical states with positive-definite distributions, e.g., Gaussian states with squeezing, which are readily accessible within current technology and provide important practical resources for CV quantum informatics \cite{Weedbrook2012}. It is fundamentally and practically important to have a simple test manifesting nonclassicality \cite{Richter2002, Mari2011} beyond the negativity in phase space.

In this respect, there was a seminal work by Banaszek and W{\'o}dkiewicz \cite{Banaszek1999} (BW), who proposed a method to test the Bell nonlocality directly in phase space. Unlike the Bell test using homodyne detection, which requires the transformation of a Gaussian state to a non-Gaussian state having a nonpositive Wigner function \cite{Bell, Nha2004}, BW formalism enables us to detect nonclassical correlation even with a positive-definite Wigner function. It has been extended to generalized quasiprobability functions \cite{Lee2009} and multipartite systems \cite{Li2011, Lee2013, Kim2013, Adesso2014}. Recently, we have theoretically proposed and experimentally demonstrated a single-mode nonclassicality test using the Wigner function \cite{Park2015} in analogy with BW formalism. 

We here extend this recent work by using generalized quasiprobability functions. We not only give more details of the proposal in \cite{Park2015}, but we also investigate other relevant aspects, e.g., robustness of our tests against experimental imperfections including photon loss, finite data, and a nonoptimal choice of phase-space points. Furthermore, we introduce an optimized test of genuine non-Gaussianity employing three phase-space points, as an addition to the four-point test in \cite{Park2015}. We also make a direct connection between our single-mode test and the BW nonlocality test, particularly showing that the single-mode nonclassical states detected under our parallelogram test can be a useful resource to make a two-mode state manifesting nonlocality under the BW test. 

Our starting point is the observation that every pair of orthogonal quadratures in phase space behaves as independent realistic variables for a coherent state. Exploiting it, we propose two nonclassicality tests that take four and three points at the vertices of a rectangle and a right triangle, respectively. Our tests detect a broad range of nonclassical Gaussian states, including all pure states. Identifying the upper bounds for all Gaussian states and their mixtures, we also propose tests for genuine quantum non-Gaussianity.
Non-Gaussian resources are known to be essential for many quantum informatic tasks, including universal CV quantum computation \cite{Lloyd1999}, entanglement distillation \cite{Eisert2002}, quantum error correction \cite{Niset2009}, and CV nonlocality testing \cite{Nha2004, Park2012}. A simple method to obtain a non-Gaussian state would be to prepare a finite mixture of Gaussian states. However, we cannot claim such a state as a genuine non-Gaussian resource. In dealing with non-Gaussianity in quantum phase space, it is important to distinguish a genuinely quantum non-Gaussian state from a mixture of Gaussian states \cite{Filip2014, Genoni2013}. We further optimize our non-Gaussianity tests by taking points from a parallelogram (triangle) instead of a rectangle (right triangle), which essentially realizes a squeezing operation on a given state without actually implementing it.

We also discuss the fundamental and the practical limits of our tests. Note that there exists a one-to-one correspondence between a $s$-parametrized phase-space distribution and a loss mechanism, i.e., interaction with a vacuum reservoir \cite{Wallentowitz1996, Lee2009}. This correspondence might suggest that the limits of the $s$-parametrized distributions are the same as those of the Wigner function ($s=0$) under a lossy Gaussian channel. However, they can yield different results for a nontrivial test. We show that the former can detect more states than the latter when a test sets a bound varying with the parameter $s$, e.g., the case of our non-Gaussianity tests. We demonstrate that our tests are able to detect a nonclassical state reliably even when the number of data is finite and the measurement setting deviates from the optimal setting. 
We also show that our parallelogram test can detect genuine non-Gaussianity for a range of superposition states with loss above 50\%, at which the Wigner function becomes positive definite.
Finally, we show that our parallelogram test can identify useful resources for a nonlocality test in phase space. It may open a direction for future works, e.g., on a deeper understanding of the relation between nonclassicality and nonlocality.

\section{Nonclassicality tests}
The $s$-parametrized quasiprobability function of a quantum state $\rho$ is defined as \cite{Barnett}
	\begin{equation}
		W_{\rho} ( q, p ; s ) = \frac{2}{\pi (1-s)} \mathrm{tr} [ \rho \hat{D}^{\dag} ( \alpha ) \hat{T} ( s ) \hat{D} ( \alpha ) ],
	\end{equation}
where $\hat{D} ( \alpha ) = \exp ( \alpha \hat{a}^{\dag} - \alpha^{*} \hat{a} )$ is the displacement operator with complex amplitude $\alpha = q + ip$. 
The operator $\hat{T} ( s )$ is given by
	\begin{equation}
		\hat{T} ( s ) \equiv \bigg( \frac{s+1}{s-1} \bigg)^{\hat{n}} = \sum_{n = 0}^{\infty} \bigg( \frac{s+1}{s-1} \bigg)^{n} \ketbra{n}{n},
	\end{equation}
which yields, e.g., a parity operator $(-1)^{\hat{n}}$ and a vacuum state $\ketbra{0}{0}$ for $s = 0$ and $-1$, corresponding to the well-known Wigner and $Q$ functions, respectively. As the eigenvalues of $\hat{T} ( s )$ are $( \frac{s+1}{s-1} )^{n}$ ($n$: non-negative integers), the $s$-parametrized quasiprobability function is bounded as
	\begin{align} \label{eq:SFAB}
		\frac{s+1}{s-1} \leq \frac{\pi (1-s)}{2} W_{\rho} ( q, p; s ) \leq 1 & \quad \mbox{for $-1 \leq s \leq 0$}, \nonumber \\
		0 < \frac{\pi (1-s)}{2} W_{\rho} ( q, p; s ) \leq 1 & \quad \mbox{for $s < -1$}.
	\end{align}
This shows that the lower bound becomes minimum for the Wigner function ($s=0$) and approaches zero with $s$ decreasing. The above equation also clearly tells us that the $Q$ function ($s=-1$) is non-negative. While the parameter $s$ can have a positive value up to 1 ($P$ function), we only deal with a nonpositive $s$ throughout the paper as the eigenvalues of $\hat{T} (s)$ become unbounded for $s > 0$.

	\begin{figure}[!t]
		\includegraphics[scale=0.4]{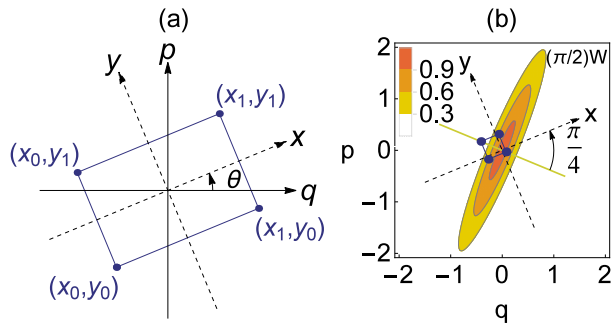}
		\caption{(Color online) (a) Nonclassicality test in Eq.~\eqref{eq:R} takes four points at the vertices of a rectangle. Neglecting a point $( x_{0}, y_{0} )$, the remaining three points form a right triangle in Eq.~\eqref{eq:RT}. (b) Optimal choice of rectangle for testing a squeezed state occurs at $\theta - \phi = \frac{\pi}{4}$ ($\phi$: squeezing axis), that is, when the axes of the rectangle are oriented midway between the squeezed and the antisqueezed axes. }
		\label{fig:rectangle}
	\end{figure}

Interestingly, the $s$-parametrized quasiprobability function of a coherent state $\ket{\alpha}$ is factorized as a product of two Gaussian distributions for every pair of orthogonal quadratures,
	\begin{align}
		& \frac{\pi (1-s)}{2} W_{\ketbra{\alpha}{\alpha}} ( x, y; s ) \nonumber \\
		& = \exp \bigg( - \frac{2 ( x - \alpha_{x} )^{2}}{1-s} \bigg) \exp \bigg( - \frac{2 ( y - \alpha_{y} )^{2}}{1-s} \bigg),
	\end{align}
where $(x,y)^{T} = \mathcal{R} ( \theta ) (q,p)^{T}$ is a coordinate system rotated by an angle $\theta$ from $(q,p)$ [Fig. 1 (a)], with a rotation matrix
	\begin{equation}
		\mathcal{R} ( \theta ) = \begin{pmatrix}
			\cos \theta & \sin \theta \\
			- \sin \theta & \cos \theta
		\end{pmatrix},
	\end{equation}
and $\alpha_{x} = \mathrm{Re} [ \alpha e^{-i \theta} ]$ and $\alpha_{y} = \mathrm{Im} [ \alpha e^{-i \theta} ]$. It is thus possible to consider the generalized quasiprobability function of a coherent state as a product of two independent random variables $a$ and $b$, that is, $\frac{\pi (1-s)}{2} W_{\ketbra{\alpha}{\alpha}} ( x, y; s ) = ab$ with $0 < a,b \leq 1$.\\

\subsection{Rectangle test}
We then construct a linear sum of $s-$parametrized functions at four phase-space points as
	\begin{align} \label{eq:R}
		\mathcal{J}_{s} [ \rho ] & \equiv \frac{\pi (1-s)}{2} \{ W_{\rho} ( x_{0}, y_{0} ; s ) + W_{\rho} ( x_{1}, y_{0} ; s ) \nonumber \\
		& + W_{\rho} ( x_{0}, y_{1} ; s ) - W_{\rho} ( x_{1}, y_{1} ; s ) \},
	\end{align}
where the points constitutes a rectangle oriented at angle $\theta$ in phase space as depicted in Fig.~\ref{fig:rectangle}. We then obtain $\mathcal{J}_{s} = a_{0} b_{0} + a_{1} b_{0} + a_{0} b_{1} - a_{1} b_{1}$ for a coherent state, which has the same form as the Clauser-Horne-Shimony-Holt (CHSH) inequality \cite{CHSH}. Using $0 < a, b \leq 1$, we obtain $-1 < \mathcal{J}_{s} \leq 2$ for a coherent state as follows. (i) For $a_{0} < a_{1}$, we have $-1 < - ( a_{1} - a_{0} ) b_{1} < \mathcal{J}_{s} < ( a_{0} + a_{1} ) b_{0} \leq 2$. The minimum is given by, e.g., $a_{0} = b_{0} = 0$ and $a_{1} = b_{1} = 1$. (ii) For $a_{0} \geq a_{1}$, we have $0 < \mathcal{J}_{s} \leq ( a_{0} + a_{1} ) + ( a_{0} - a_{1} ) \leq 2$. The maximum is given by, e.g., $a_{0} = b_{0} = a_{1} = b_{1} = 1$.

We extend the above result to an arbitrary classical state, i.e. a mixture of coherent states, $\rho_{\mathrm{cl}} = \int d \lambda p(\lambda) \ketbra{\lambda}{\lambda}$, where $\ket{\lambda}$ is a coherent state. As the coherent amplitude $\lambda$ behaves like a hidden variable,
	\begin{equation} \label{eq:HV}
		\frac{\pi (1-s)}{2} W_{\rho_{\mathrm{cl}}} ( x, y; s ) = \int d \lambda p ( \lambda ) a(x|\lambda) b(y|\lambda),
	\end{equation}
we obtain a classicality condition as
	\begin{equation} \label{eq:RC}
		-1 < \mathcal{J}_{s} [ \rho_{\mathrm{cl}} ] \leq 2.
	\end{equation}
In other words, the violation of Eq.~\eqref{eq:RC} demonstrates the nonclassicality of a single-mode state.

At this point, it may be intriguing to ask how many phase-space points should be considered to come up with a meaningful nonclassicality test, particularly to test a positive quasiprobability distribution. Can we obtain a useful nonclassicality criterion employing less numbers of points than four in Eq.~\eqref{eq:R}? Of course, verifying nonclassicality from an {\it arbitrary} set of points is impossible without specifications on the chosen points $\{ q_{j}, p_{j} \}$, like a predetermined position (origin) with energy constraint in \cite{Genoni2013} or a designated shape (rectangle) in our case \cite{Park2015}. 
For example, if we construct a test exploiting the values at fully arbitrary $N$ points without specifying locations, that is, $\mathcal{F} [ v_{1}, ..., v_{N} ]$ where $v_{j} = \frac{\pi (1-s)}{2} W_{\rho} ( q_{j}, p_{j}; s )$ with $j \in \{ 1, ..., N \}$, every result from a set of positive values ($v_{1}, ..., v_{N}$) is mimicked by a single vacuum state because the same values can be found at $( q_{j}^{\prime}, p_{j}^{\prime} ) = ( 0, \sqrt{\frac{s-1}{2} \log v_{j}}, s )$ for $j \in \{ 1, ..., N \}$. In this sense, if we intend to introduce a specified shape as a constraint, the least number of points is possibly three with the shape of triangle, whereas the test in Eq.~\eqref{eq:R} adopts a rectangle with four points.\\

\subsection{Right-triangle test}
Thus we also introduce a three-points (right triangle) test as
	\begin{align} \label{eq:RT}
		\mathcal{J}_{s}^{\prime} [ \rho ] & \equiv \frac{\pi (1-s)}{2} \{ W_{s} ( x_{1}, y_{0} ) + W_{s} ( x_{0}, y_{1} ) \nonumber \\
		& - W_{s} ( x_{1}, y_{1} ) \},
	\end{align}
which excludes one point $( x_{0}, y_{0} )$ from Eq.~\eqref{eq:R}. We then have a structure $\mathcal{J}_{s}^{\prime} = a_{1}b_{0} + a_{0}b_{1} - a_{1}b_{1}$ for a coherent state. Using $0 < a,b \leq 1$ again, we obtain $-1 < \mathcal{J}_{s}^{\prime} \leq 1$ as follows: (i) For $a_{0} < a_{1}$, we have $-1 < - ( a_{1} - a_{0} ) b_{1} < \mathcal{J}_{s}^{\prime} < a_{1} b_{0} \leq 1$. The minimum is achieved by $a_{0} = b_{0} = 0$ and $a_{1} = b_{1} = 1$. (ii) For $a_{0} \geq a_{1}$, we have $0 < \mathcal{J}_{s}^{\prime} < a_{1} + ( a_{0} - a_{1} ) \leq 1$. The maximum is achieved by $a_{0} = b_{0} = a_{1} = b_{1} = 1$.

Therefore, similar to Eq.~\eqref{eq:HV}, we obtain another classicality condition as
	\begin{equation} \label{eq:RTC}
		-1 < \mathcal{J}_{s}^{\prime} [ \rho_{\mathrm{cl}} ] \leq 1.
	\end{equation}
We note that, contrary to the rectangle test in Eq.~\eqref{eq:R}, it has no analogy with a nonlocality test, as $a_{1}b_{0} + a_{0}b_{1} - a_{1}b_{1}$ is saturated by a hidden variable theory: $a_{0} = b_{0} = - a_{1} = b_{1} = \pm 1$ yields $\pm 3$, which are also the quantum bounds due to $\left|\frac{\pi (1-s)}{2} W_{s} ( x, y)\right|\le1$.

\subsubsection{Invariance under displacement and phase-rotation operations}
Note that the optimal values of $\mathcal{J}_{s} [ \rho ]$ and $\mathcal{J}_{s}^{\prime} [ \rho ]$ for a given state $\rho$ are invariant under displacement and rotation. Let us assume that a state $\rho$ has an optimal value at points $\{ x_{0}, y_{0}, x_{1}, y_{1} \}$, and then a displaced state $\hat{D} ( \alpha ) \rho \hat{D}^{\dag} ( \alpha )$ has the same optimum at shifted points $\{ x_{0} + \alpha_{x}, y_{0} + \alpha_{y}, x_{1} + \alpha_{x}, y_{1} + \alpha_{y} \}$. This is because the displacement operator only translates the center of the quasiprobability function while preserving its entire profile. Similarly, if a state has the optimal value at points $\{ x_{0}, y_{0}, x_{1}, y_{1} \}$ where the coordinate system is oriented at angle $\theta$, a rotated state $e^{i \varphi \hat{a}^{\dag} \hat{a}} \rho e^{- i \varphi \hat{a}^{\dag} \hat{a}}$ has the same optimum in the coordinate system now oriented at angle $\theta + \varphi$, since the phase-rotation also preserves the profile of the quasiprobability function. 
These invariance properties can be useful to simplify the analysis of nonclassicality tests for a given state.

\subsection{Gaussian states}
We first demonstrate how our tests $\mathcal{J}_{s}$ and $\mathcal{J}_{s}^{\prime}$ can detect a wide range of Gaussian states. A single-mode Gaussian state $\sigma$ is fully characterized by its first-order moments (averages) $\expect{\hat{q}}$ and $\expect{\hat{p}}$, and second-order moments represented by a covariance matrix $\Gamma$ with elements
	\begin{equation}
		\Gamma_{jk} = \frac{1}{2} \expect{\hat{Q}_{j} \hat{Q}_{k} + \hat{Q}_{k} \hat{Q}_{j}} - \expect{\hat{Q}_{j}} \expect{\hat{Q}_{k}}, \quad (j,k=1,2)
	\end{equation}
where $\hat{Q} \equiv ( \hat{q}, \hat{p} )^{T}$ with $\hat{q} = \frac{1}{2} ( \hat{a} + \hat{a}^{\dag} )$ and $\hat{p} = \frac{1}{2i} ( \hat{a} - \hat{a}^{\dag} )$. Its $s$-parametrized distribution is given by a Gaussian function as
	\begin{align} \label{eq:SG}
		& W_{\sigma} ( q, p; s ) \nonumber \\
		& = \frac{2 f_{s}}{\pi (1-s)} \exp \bigg[ - \frac{1}{2} ( Q - \expect{Q} )^{T} \Gamma_{s}^{-1} ( Q - \expect{Q} ) \bigg],
	\end{align}
where $Q = (q,p)^{T}$, $f_{s} = \frac{1-s}{4 \sqrt{\det \Gamma_{s}}}$ and $\Gamma_{s} = \Gamma - \frac{s}{4} \mathbb{I}$.

	\begin{figure*}[!t]
		\includegraphics[scale=0.5]{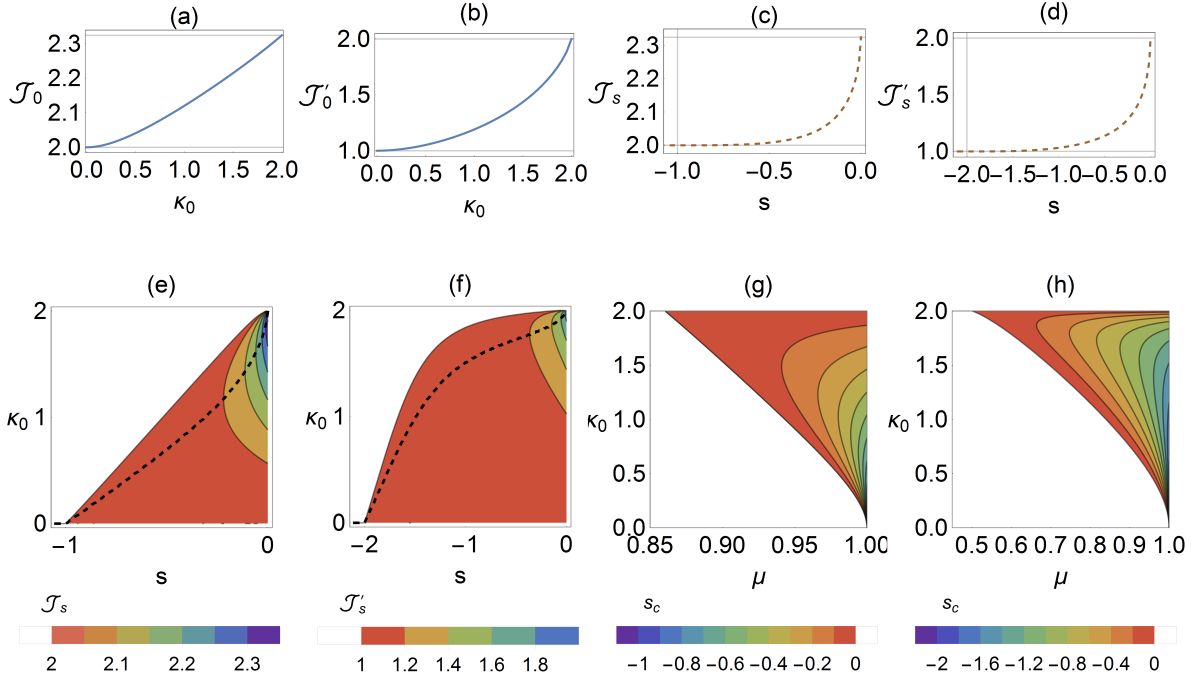}
		\caption{(Color online) (a, b) Maximum $\mathcal{J}_{s=0}$ and $\mathcal{J}_{s=0}^{\prime}$ (test with Wigner function $s=0$) for Gaussian states with respect to squeezing strength $\kappa_{0} = 2 \tanh 2r$. (c, d) Maximum values of $\mathcal{J}_{s}$ and $\mathcal{J}_{s}^{\prime}$, respectively, among all Gaussian states with respect to $s$ (generalized quasidistributions). These maximum values become critical  bounds to test genuine non-Gaussianity for each $s$ in Eq. (35). (e, f) Contour plots for optimal $\mathcal{J}_{s}$ and $\mathcal{J}_{s}^{\prime}$ with respect to $s$ and $\kappa_{0}$. Black dashed lines represent the squeezing strength $\kappa_{0}$ that yields the maximal values shown in (c) and (d). (g, h) Contour plots of a critical parameter $s_c$ for a Gaussian state with purity $\mu$ and squeezing strength $\kappa_{0}$, above which nonclassicality can be detected via rectangle and triangle testing, respectively. The colored regions thus represent the parameter space of Gaussian states in which there exist $s$-parametrized distributions $s\in[s_c,0]$ for a successful nonclassicality test.}
		\label{fig:Gaussian}
	\end{figure*}

Alternatively, a single-mode Gaussian state $\sigma$ can be represented as a displaced squeezed thermal state,
	\begin{equation} \label{eq:DSTS}
		\sigma = \hat{D} ( \alpha ) \hat{S} ( r, \phi ) \sigma_{th} ( \bar{n} ) \hat{S}^{\dag} ( r, \phi ) \hat{D}^{\dag} ( \alpha ),
	\end{equation}
where $\hat{S} ( r, \phi ) = \exp [ - \frac{r}{2} ( e^{2i\phi} \hat{a}^{\dag 2} - e^{-2i\phi} \hat{a}^{2} ) ]$ is the squeezing operator ($r$: squeezing strength, $\phi$: squeezing axis), and $\sigma_{th} ( \bar{n} ) = \sum_{n = 0}^{\infty} \frac{\bar{n}^{n}}{( \bar{n} + 1 )^{n+1}} \ketbra{n}{n}$ is a thermal state with mean photon number $\bar{n}$.  For a Gaussian state with parameters $\{ \alpha, r, \phi, \bar{n} \}$, its first moments are given by $\expect{\hat{q}} = \mathrm{Re} [ \alpha ]$ and $\expect{\hat{p}} = \mathrm{Im} [ \alpha ]$, and the covariance matrix elements by
	\begin{align}
		\Gamma_{11} & = \frac{1}{2} \bigg( \bar{n} + \frac{1}{2} \bigg) ( \cosh 2r - \sinh 2r \cos 2\phi ), \nonumber \\
		\Gamma_{22} & = \frac{1}{2} \bigg( \bar{n} + \frac{1}{2} \bigg) ( \cosh 2r + \sinh 2r \cos 2\phi ), \nonumber \\
		\Gamma_{12} & = \Gamma_{21} = - \frac{1}{2} \bigg( \bar{n} + \frac{1}{2} \bigg) \sinh 2r \sin 2\phi,
	\end{align}
which yields
	\begin{equation} \label{eq:OF}
		f_{s} = \frac{1-s}{\sqrt{(1+2\bar{n})^{2}+s^{2}-2(1+2\bar{n})s\cosh 2r}}.
	\end{equation}
Equation~\eqref{eq:OF} shows that the overall factor of $W_{\sigma} ( q, p ; s )$ in Eq. (12) is bounded by $0 < f_{s} \leq 1$. Its maximum $f_s=1$ is achieved by every pure Gaussian state ($\bar{n}=0$) for $s=0$, as $f_{s=0}=\frac{1}{1+2\bar{n}}$ represents the purity of a Gaussian state $\mu \equiv \mathrm{tr} \sigma^{2} = \frac{1}{1+2\bar{n}}$. On the other hand, only a vacuum state attains the maximum $f_s=1$ for $s < 0$. In general, with $r$ and $s$ fixed, $f_{s}$ increases with purity ($\bar{n}$ decreasing).

Rewriting Eq.~\eqref{eq:SG} using the rotated quadratures $\widetilde{Q} = (x,y)^{T} = \mathcal{R}(\theta) (q,p)^{T}$, we obtain
	\begin{align} \label{eq:RSG}
		& W_{\sigma} ( x, y; s ) \nonumber \\
		& = \frac{2 f_{s}}{\pi (1-s)} \exp \bigg[ - \frac{1}{2} ( \widetilde{Q} - \expect{\widetilde{Q}} )^{T} \widetilde{\Gamma}_{s}^{-1} ( \widetilde{Q} - \expect{\widetilde{Q}} ) \bigg],
	\end{align}
where $\widetilde{\Gamma}_{s} = \mathcal{R}(\theta)\Gamma_{s}\mathcal{R}(-\theta)$ is the covariance matrix in a rotated frame and $\expect{\widetilde{Q}} = ( \mathrm{Re} [ \alpha e^{-i \theta} ], \mathrm{Im} [ \alpha e^{-i \theta} ] )$.
From now on, we set $\alpha = 0$, as the displacement operation has no effect on the optimal values (Sec. II~B~1). Every Gaussian function in the form of Eq.~\eqref{eq:RSG} can be recast to
	\begin{equation}
		\frac{\pi (1-s)}{2} W_{\sigma} ( x, y; s ) = f_{s} \exp [ - \widetilde{x}^{2} - \widetilde{y}^{2} + k_{s} \widetilde{x} \widetilde{y} ],
	\end{equation}
by introducing rescaled variables,
	\begin{align}
		( \widetilde{x}, \widetilde{y} ) = \frac{2 \sqrt{2} f_{s}}{1-s} \bigg( \sqrt{\widetilde{\Gamma}_{22} - \frac{s}{4}} x, \sqrt{\widetilde{\Gamma}_{11} - \frac{s}{4}} y \bigg),
	\end{align}
and the parameter
	\begin{align}
		k_{s} & = \frac{2 \widetilde{\Gamma}_{12}}{\sqrt{( \widetilde{\Gamma}_{11} - \frac{s}{4} ) ( \widetilde{\Gamma}_{22} - \frac{s}{4} )}} \nonumber \\
		& = \frac{2 \sinh 2r \sin 2 ( \theta - \phi )}{\sqrt{( \frac{s}{1+2n} - \cosh 2r )^{2} - \sinh^{2} 2r \cos^{2} 2 ( \theta - \phi )}}.
	\end{align}
From the rescaled distributions in Eq. (17), we see that two parameters, i.e. the overall factor $f_s$ and the parameter $k_s$, determine optimal values $\mathcal{J}_{s}$ and $\mathcal{J}_{s}^{\prime}$ for a given Gaussian state.
In fact, we can show that these optimal values monotonically increase with $k_s$ as well as $f_s$ \cite{Park2015}.
We point out that the parameter $k_s$ for every Gaussian state is bounded by $-2 \leq k_{s} \leq 2$. 
In particular, with $s$ and $n$ fixed, the parameter $k_s$ is bounded by 
	\begin{equation}
		|k_{s}| \leq \frac{2 \sinh 2r}{- \frac{s}{1+2n} + \cosh 2r} \equiv \kappa_{s},
	\end{equation}
where the upper bound $\kappa_{s}$ is obtained at the choice of angle $\theta - \phi = \frac{\pi}{4}$ (see Fig.~\ref{fig:rectangle}). This optimal choice of angle intuitively makes sense, as our nonclassicality tests rely on the {\it nonfactorizability} of the quasi-probability distributions in phase space. In contrast, if we take $\theta - \phi = 0$ or $\frac{\pi}{2}$, the quasiprobability function in Eq.~\eqref{eq:RSG} is factorized to a form $W_{\sigma} ( x, y; s ) = W ( x ) W ( y )$, where the two quadratures $x$ and $y$ behave as independent variables, yielding no violation of our nonclassicality tests.\\

\subsubsection{Maximum values of $\mathcal{J}_{s}$ and $\mathcal{J}_{s}^{\prime}$ for Gaussian states}

For $s=0$ (Wigner function), the overall factor $f_{0} = \frac{1}{1+2 \bar{n}}$ depends only on purity; thus large $\mathcal{J}_{0}$ and $\mathcal{J}_{0}^{\prime}$ occur for a pure state. On the other hand, the optimal $\kappa_{s=0} = 2 \tanh 2r$ in Eq. (20) depend only on the degree of squeezing. In Figs.~\ref{fig:Gaussian}(a) and~\ref{fig:Gaussian}(b), we show that the optimal $\mathcal{J}_{0}$ and $\mathcal{J}_{0}^{\prime}$ monotonically increase with $\kappa_{0}$ (degree of squeezing). They rise up to the maximum values $\mathcal{J}_{0} = \frac{8}{3^{9/8}} \approx 2.32$ and $\mathcal{J}_{0}^{\prime} = 2$ for the rectangle and the right triangle tests, respectively, both achieved at an infinite squeezing $\kappa_{0} = 2$ ($r \rightarrow \infty$). (See Supplemental Matrerial of Ref. \cite{Park2015} for rigorous proofs of the maximal values.) Note that the Gaussian bound $\frac{8}{3^{9/8}}$ for the four-points test coincides with the maximal value of a Gaussian state for two-mode \cite{Jeong2003} and three-mode \cite{Adesso2014} nonlocality tests in phase space.

On the other hand, for $s < 0$, the overall factor $f_{s}$ and the ratio $\kappa_{s}$ involve both the purity $\mu$ and the squeezing strength $r$. With purity and squeezing fixed, both $f_{s}$ and $\kappa_{s}$ decrease with the parameter $s$ decreasing. We thus observe that maximum values among all Gaussian states for rectangle and right triangle tests, respectively, become smaller with the parameter $s$ decreasing in Fig.~\ref{fig:Gaussian}(c) and~\ref{fig:Gaussian}(d), respectively. 
In addition, although the optimal $\kappa_s$ in Eq. (20) increases with the thermal photon $\bar{n}$ for $s<0$, the overall factor $f_s$ in Eq. (15) decreases with $\bar{n}$, which eventually makes the case of a pure state ($\bar{n}=0$) optimal for given $r$ and $s$. 

We plot the optimal values of $\mathcal{J}_{s}$ [Fig.~\ref{fig:Gaussian} (e)] and $\mathcal{J}_{s}^{\prime}$ [Fig.~\ref{fig:Gaussian} (f)] with respect to parameter $s$ and squeezing strength $\kappa_{0} = 2 \tanh 2r$. 
In these contour plots, we show how the optimal squeezing $\kappa_{0}$ for each maximum changes with $s$ (black dashed lines). Interestingly, we note that the optimal violation for a nonzero $s<0$ occurs at a finite squeezing, similar to the case of a two-mode nonlocality test \cite{Lee2009}. With the parameter $s$ decreasing, the maximum values and the corresponding squeezing strength become smaller. At $s=-1$ and $s=-2$, there is no Gaussian state violating the rectangle test and the triangle test, respectively, which suggests that the right-triangle test is more useful practically for detecting Gaussian states. 

In Figs.~\ref{fig:Gaussian}(g) and~\ref{fig:Gaussian}(h), we also identify the range of mixed Gaussian states (colored region) that can be detected under our nonclassicality tests. Specifically, we plot the critical parameter $s_c$ for each Gaussian state with purity $\mu$ and squeezing $\kappa_0$, above which its nonclassicality can successfully be detected, i.e., in the range $s\in[s_c,0]$. As the purity $\mu$ decreases, we see that the squeezing level $\kappa_0$ required for a successful test becomes higher. In addition, the two contour plots (g) and (h) in comparison show that right triangle test detects more Gaussian states than the rectangle test. 
The range of successful detection may be attributed to the ratio of maximum Gaussian bound to classical bound in each test. The rectangle test gives the ratio $\left(\frac{8}{3^{9/8}}\right)/2\approx 1.16$, whereas the right-triangle test gives $2/1=2$, which may account for the resilience of the latter test compared to the former. For instance, violating $\mathcal{J}_{0} \geq 2$ and $\mathcal{J}_{0}^{\prime} \geq 1$ becomes impossible if the purity falls below the inverse of the raios, $\frac{3^{9/8}}{4} \approx 0.86$ and $\frac{1}{2}$, respectively, which has been numerically confirmed.\\

\subsubsection{s-parametrized functions and loss mechanism}

In addition, Fig.~\ref{fig:Gaussian} shows that Wigner function ($s=0$) is optimal among all $s$-parametrized distributions for both tests, which makes sense as the $s$-parametrized quasiprobability function is closely related to loss dynamics under a Gaussian reservoir. The $s$-parametrized quasiprobability functions of a quantum state $\rho$ with two different parameters $s_{1}$ and $s_{2}$ $(s_{2}<s_{1})$ are related by a Gaussian convolution \cite{Barnett},
	\begin{align} \label{eq:GC}
		& W_{\rho} ( \alpha;  s_{2} ) \nonumber \\
		& = \frac{2}{\pi (s_{1}-s_{2})}  \int d^{2} \beta W_{\rho} ( \beta, s_{1} )  e^{- \frac{2}{s_{1} - s_{2}} | \alpha - \beta |^{2}}.
	\end{align}
Equation~\eqref{eq:GC} is similar to the action of a loss channel $\mathcal{L}$ on a state $\rho$,
	\begin{align}
		& W_{\mathcal{L} [ \rho ]} ( \alpha; 0 ) \nonumber \\
		& = \frac{2}{\pi ( 1 - \eta )} \int d^{2} \beta W_{\rho} ( \beta; 0 ) e^{- \frac{2}{1 - \eta} | \alpha - \sqrt{\eta} \beta |^{2}},
	\end{align}
where the loss channel $\mathcal{L}$ is modeled by mixing the input state $\rho$ and a vacuum at a beam splitter with transmittance $\eta$. It provides a direct connection between the generalized quasiprobability function and loss dynamics as \cite{Wallentowitz1996, Lee2009}
	\begin{equation} \label{eq:VR}
		W_{\mathcal{L} [ \rho ]} ( \alpha; s ) = \frac{1}{\eta} W_{\rho} \bigg( \frac{\alpha}{\sqrt{\eta}}; 1 - \frac{1-s}{\eta} \bigg).
	\end{equation}
Inverting Eq.~\eqref{eq:VR} by setting $s = 0$ and $s^{\prime} = 1 - \frac{1}{\eta}$, we obtain
	\begin{equation} \label{eq:VRF}
		\frac{\pi (1-s^{\prime})}{2} W_{\rho} ( \sqrt{1-s^\prime} \alpha; s^{\prime} ) =  \frac{\pi}{2} W_{\mathcal{L} [ \rho ]} ( \alpha; 0 ),
	\end{equation}
which reveals that $\mathcal{J}_{s}$ and $\mathcal{J}_{s}^{\prime}$ in Eqs.~\eqref{eq:R} and \eqref{eq:RT} correspond to loss dynamics of $\mathcal{J}_{0}$ and $\mathcal{J}_{0}^{\prime}$, respectively, with $\eta = \frac{1}{1-s^{\prime}}$. The results in Fig.~\ref{fig:Gaussian} thus identify the ultimate limit of our tests for Gaussian states under a lossy channel. That is, as the rectangle and the triangle tests for Gaussian states have critical values $s_c=-1$ and $s_c=-2$, respectively, we have 3 dB ($\eta_c = \frac{1}{1-s_c}=\frac{1}{2}$) and 4.77 dB ($\eta_c =\frac{1}{3}$) loss limits, below which there exist some Gaussian states detectable using our tests. 
The different limits 3 dB and 4.77 dB manifest a practical superiority of the triangle test to the rectangle test.\\


\subsubsection{finite data and nonoptimal phase-space points}

Now let us further investigate to what extent our tests can be useful under practical conditions. First, the number $N$ of data to construct an average value is always finite, incurring an error of order $O(\frac{1}{\sqrt{N}})$. We thus require that the degree of violation is large enough to overcome the statistical error as
	\begin{equation}\label{eq:stat}
		\expect{\mathcal{J}} > B_c + \frac{\Delta{\mathcal{J}}}{\sqrt{N}},
	\end{equation}
where $\expect{\mathcal{J}^k} = \frac{1}{N} \sum_{i=1}^{N} \mathcal{J}_{i}^k$ ($k=1,2$) and $\Delta^2{\mathcal{J}} = \expect{\hat{\mathcal{J}}^{2}} - \expect{\hat{\mathcal{J}}}^{2}$ with the classicality bound $B_c=2$ and 1 for the rectangle and the right-triangle test, respectively. Moreover, although we have previously obtained the optimal choice of angle $\theta - \phi= \frac{\pi}{4}$ as shown in Fig. 1, it is of practical interest to identify the angle tolerance $\Delta$, i.e., a successful detection in the range of angles $|\theta - \phi -\frac{\pi}{4}|\leq\frac{\Delta}{2}$. This is a particularly important issue when there is no information on the phase (squeezing angle $\phi$) of the state. In this case, our choice of angle $\theta$ for a rectangle (triangle) in Fig. 1 becomes completely random. 
A worst case would be the choice of $\theta - \phi=0$ or $\frac{\pi}{2}$ at which no violation occurs due to the factorizability of the phase-space function, as explained below Eq. (20).

\begin{figure}
		\includegraphics[width=\linewidth]{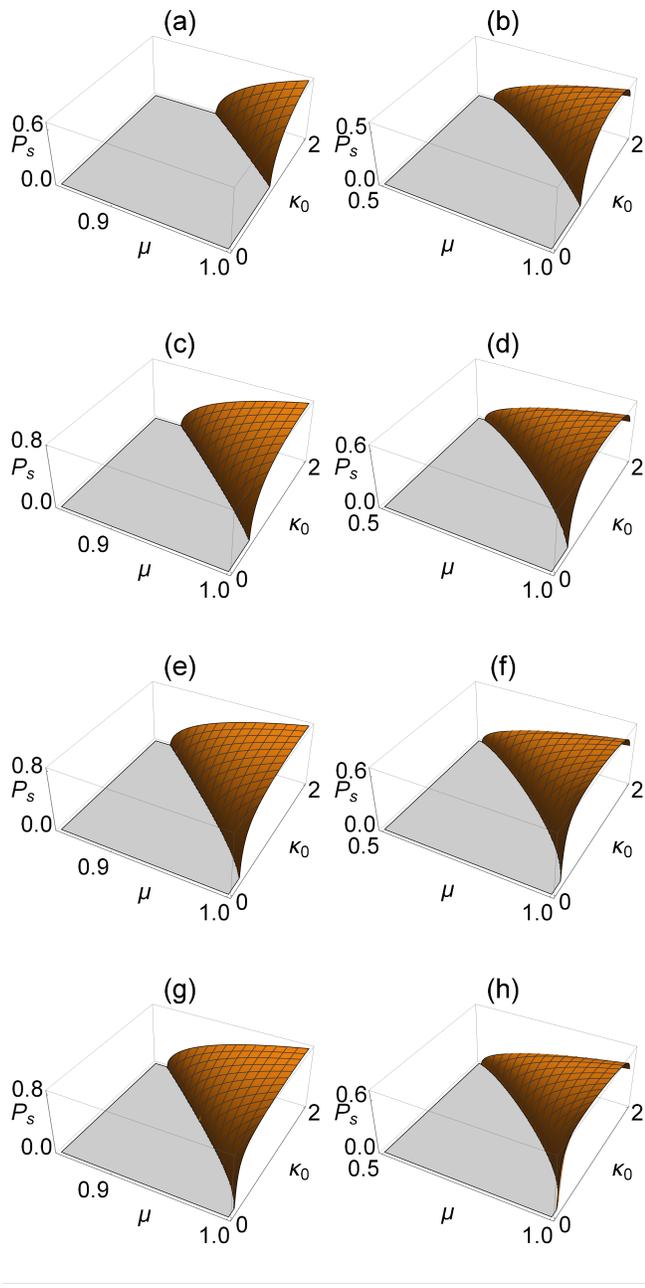}
		\caption{(Color online) Success probability $P_s=\Delta/(\frac{\pi}{2})$ for a rectangle test $\mathcal{J}_{s=0}$ [(a), (c), (e), and (g)] and a triangle test $\mathcal{J}_{s=0}^{\prime}$ [(b), (d), (f), and (h)] to detect the nonclassicality of a Gaussian state with purity $\mu$ and squeezing strength $\kappa$ when the phase is unknown. We have used the number of data $N$ in Eq.~\eqref{eq:stat} as $N=10^3$ (a, b), $N=10^4$ (c, d), $N=10^5$ (e, f), and $N=10^6$ (g, h).}
		\label{fig:FGAD}
	\end{figure}

For given purity $\mu$ and squeezing $\kappa_0$, we may take a fixed dimension of rectangle (triangle), like the one used for an optimal test with known phase, but consider the angle $\theta$ randomly distributed over the whole range of $\frac{\pi}{2}$. We can then measure a success probability as $P_s=\Delta/(\frac{\pi}{2})$.  
In Fig.~\ref{fig:FGAD}, we show how faithfully our Wigner-function tests $\mathcal{J}_{s=0}$ in (a,c,e,g) and $\mathcal{J}_{s=0}^{\prime}$ in (b,d,f,h) detect Gaussian states with unknown phase by evaluating $P_s$ for a data number $N=10^3$ (a,b), $N=10^4$ (c,d), $N=10^5$ (e,f), and $N=10^6$ (g,h). 
We see that our tests can confidently detect a range of mixed squeezed states with a practical number $N=10^3\sim10^6$. In general, the angle tolerance $\Delta$, and thus the success probability $P_s$, becomes large by increasing the data number $N$ as well as the purity $\mu$ and squeezing strength $\kappa$. As already identified in Fig. 2(g) and 2(h), the theoretical limits of purity for a successful test of $\mathcal{J}$ and $\mathcal{J}^{\prime}$ are $\mu\approx0.83$ and $\mu=0.5$, respectively, which are achievable with $N$ growing. In Fig.~\ref{fig:FGAD}, we already have similar levels of critical purity $\mu=0.867$ (e) and 0.516 (f) with a finite data $N=10^5$. 
For the rectangle test, a high level $P_s\sim0.8$ means that our test can be successful unless the randomly chosen angle is too close to the one for the factorized Wigner function, $\theta - \phi=0$ or $\frac{\pi}{2}$. For the triangle test, the success probability $P_s$ is smaller, however, the range of mixed states detectable is larger than that of the rectangle test.

\begin{figure}
		\includegraphics[width=\linewidth]{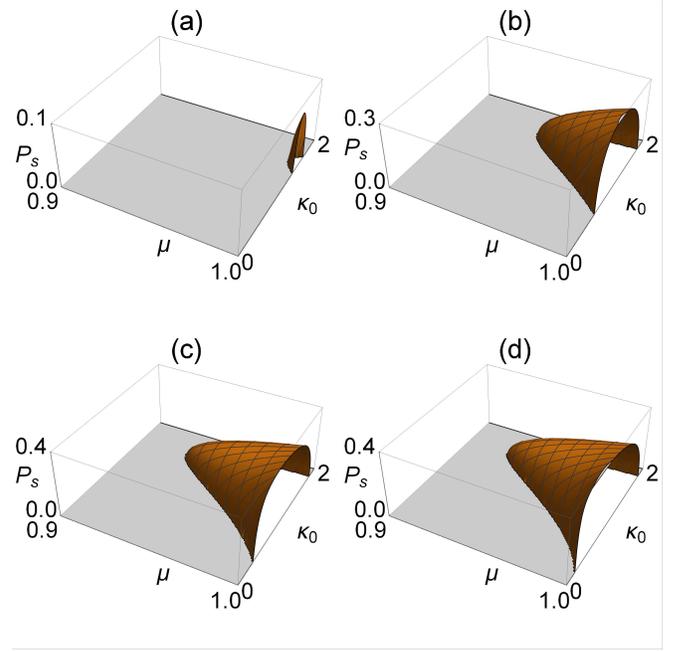}
		\caption{(Color online) Success probability $P_s=\Delta/(\frac{\pi}{2})$ for a triangle test $\mathcal{J}_{s=-1}^{\prime}$ to detect the nonclassicality of a Gaussian state with purity $\mu$ and squeezing strength $\kappa$ when the phase is unknown. We have used the number of data $N$ in Eq.~\eqref{eq:stat} as $N=10^4$ (a), $N=10^5$ (b), $N=10^6$ (c), and $N=10^7$ (d). }
		\label{fig:FGADQ}
	\end{figure}

In Fig.~\ref{fig:FGADQ}, we also show the results for a triangle test $\mathcal{J}^{\prime}_{-1}$ based on the $Q$ function ($s=-1$). Note that we have excluded the rectangle test $\mathcal{J}_{-1}$, since no violation occurs as shown in Fig. 2. Quite naturally (see Sec. II~C~2), both the detectable range of mixed Gaussian states and the success probability significantly shrink compared to the case of Wigner-function tests.

\subsection{Quantum bound beyond Gaussian states}
We now go beyond the Gaussian regime and investigate a maximal possible value of ${\mathcal{J}}$ among all quantum states beyond Gaussian states. To find out the maximum and the minimum values, we solve eigenvalue equations $\mathcal{H} \ket{\psi} = \lambda \ket{\psi}$ for Hermitian operators $\mathcal{H}_{s}$ and $\mathcal{H}_{s}^{\prime}$ that correspond to rectangle and right triangle tests, respectively. These Hermitian operators are given by
	\begin{align} \label{eq:H1}
		\mathcal{H}_{s} & \equiv \hat{D} ( q_{0} + i p_{0} ) \hat{T} ( s ) \hat{D}^{\dag} ( q_{0} + i p_{0} ) \nonumber \\
		& + \hat{D} ( q_{0} + i p_{1} ) \hat{T} ( s ) \hat{D}^{\dag} ( q_{0} + i p_{1} ) \nonumber \\
		& + \hat{D} ( q_{1} + i p_{0} ) \hat{T} ( s ) \hat{D}^{\dag} ( q_{1} + i p_{0} ) \nonumber \\
		& - \hat{D} ( q_{1} + i p_{1} ) \hat{T} ( s ) \hat{D}^{\dag} ( q_{1} + i p_{1} ),
	\end{align}
and
	\begin{align} \label{eq:H2}
		\mathcal{H}_{s}^{\prime} & \equiv \hat{D} ( q_{0} + i p_{1} ) \hat{T} ( s ) \hat{D}^{\dag} ( q_{0} + i p_{1} ) \nonumber \\
		& + \hat{D} ( q_{1} + i p_{0} ) \hat{T} ( s ) \hat{D}^{\dag} ( q_{1} + i p_{0} ) \nonumber \\
		& - \hat{D} ( q_{1} + i p_{1} ) \hat{T} ( s ) \hat{D}^{\dag} ( q_{1} + i p_{1} ).
	\end{align}
From now on, without loss of generality, we set $\{ q_{0}, p_{0}, q_{1}, p_{1} \} = \{ 0, 0, d_{q}, d_{p} \}$ since the optimal values in our tests are invariant under displacement as mentioned in Sec. II B 1. 

	\begin{figure}[!t]
		\includegraphics[scale=0.45]{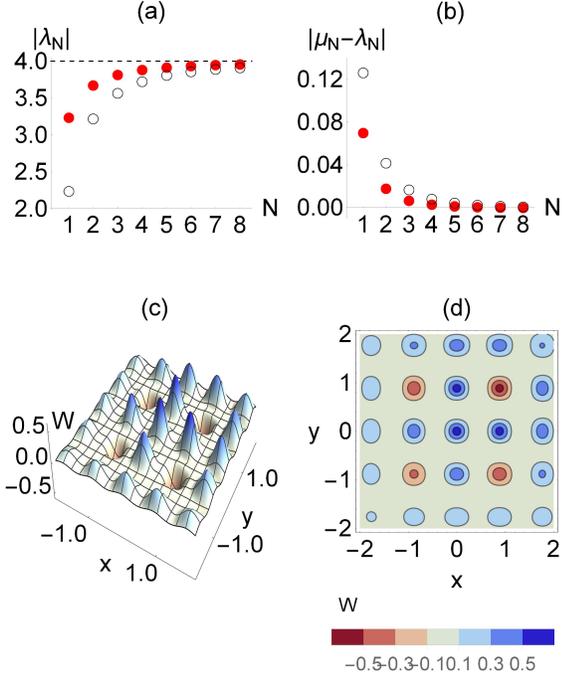}
		\caption{(Color online) (a) Maximum positive (red filled circle) and minimum negative (black open circle) eigenvalues $\lambda_{N}$ of $\mathcal{H}_{0}$ in Eq.~\eqref{eq:H1} that are obtained by truncating $n$ and $m$ up to $N$ in the recurrence relation of Eq.~\eqref{eq:recurrence}. (b) Expectation values $\mu_{N} = \expect{\mathcal{H}_{0}}$ of a truncated superposition of $(2N+1) \times (2N+1)$ coherent states in Eq.~\eqref{eq:truncation} compared with $\lambda_{N}$ in (a). As an example, we plot (c) the Wigner function and (d) its contour plot for a superposition of $5 \times 5$ coherent states with $d_{q} = d_{p} = \frac{\sqrt{\pi}}{2}$, which achieves a value of $\mathcal{J}_{0} [ \rho ] \approx 3.70$.}
		\label{fig:QB0}
	\end{figure}

For the case of the Wigner function ($s=0$), we can solve the eigenvalue problem on the basis of coherent states. We first construct a trial solution as
	\begin{equation} \label{eq:trial}
		\ket{\psi} = \sum_{n = - \infty}^{\infty} \sum_{m = - \infty}^{\infty} C_{n, m} \ket{2 d_{q} n + 2 i d_{p} m},
	\end{equation}
where the coherent states $\ket{2 d_{q} n + 2 i d_{p} m}$ with integers $n$ and $m$ form a two-dimensional lattice with points spaced by $2d_{q}$ and $2d_{p}$ along the $q$ and $p$ axes, respectively. Exploiting an identity $\hat{D} ( \alpha ) \hat{T} ( 0 ) \hat{D}^{\dag} ( \alpha ) \ket{\gamma} = e^{- \alpha \gamma^{*} + \alpha^{*} \gamma} \ket{2 \alpha - \gamma}$, we derive a recurrence relation
	\begin{align} \label{eq:recurrence}
		\lambda C_{n, m} & = C_{-n,-m} + e^{-4id^{2}m} C_{-n+1,-m} + e^{4id^{2}n} C_{-n,-m+1} \nonumber \\
		& - e^{-4id^{2} ( m - n )} C_{-n+1,-m+1},
	\end{align}
where $d^{2} \equiv d_{q} d_{p}$ refers to the area of unit cell in the lattice. In Fig.~\ref{fig:QB0}, we show the maximum and minimum eigenvalues attainable by taking a truncation number $N$ in the recurrence relations with $d^{2} = \frac{R \pi}{2} + \frac{\pi}{4}$ ($R$: integer). It shows that the simple algebraic bounds $\pm 4$ for a rectangle test are asymptotically obtained by increasing $N$. 
As a double check, we compare $\lambda_{N}$ and $\mu_{N} \equiv \bra{\psi^{N}} \mathcal{H}_{0} \ket{\psi^{N}} / \braket{\psi^{N}}{\psi^{N}}$ with $d_{q}^{2} = d_{p}^{2} = \frac{\pi}{4}$, where
	\begin{equation} \label{eq:truncation}
		\ket{\psi^{N}} = \sum_{n=-N}^{N} \sum_{m=-N}^{N} C_{n,m} \ket{2 d_{q} n + 2 i d_{p} m}
	\end{equation}
is given by plugging the coefficients $C_{n,m}$ obtained from the recurrence relations. That is, we construct a state with certain numeric coefficients and caluclate its actual average value $ \mu_{N}$ of $\mathcal{H}_{s=0}$.  Figure~\ref{fig:QB0} confirms that the algebraic bounds are actually obtained by the states in Eq.~\eqref{eq:truncation}. In addition, those states also achieve the algebraic bounds $\pm 3$ for a right triangle test (Fig. 6).

\begin{figure}[!t]
		\includegraphics[scale=0.45]{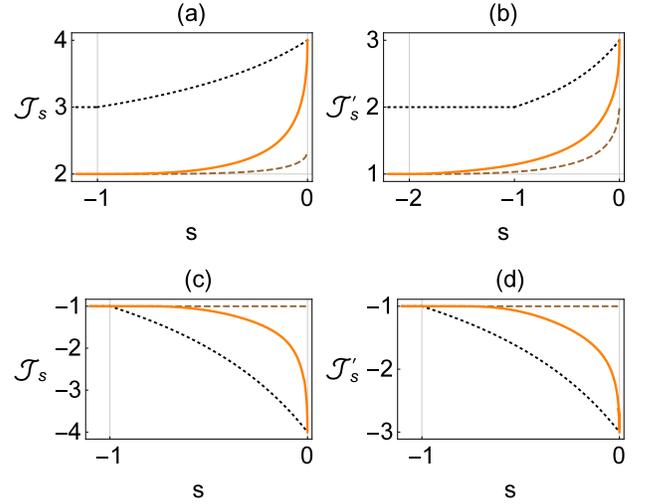}
		\caption{(Color online) (a, b) Maximal $\mathcal{J}_{s}$ ($\mathcal{J}_{s}^{\prime}$) and (c, d) minimal $\mathcal{J}_{s}$ ($\mathcal{J}_{s}^{\prime}$) among all quantum states against the parameter $s$ (orange solid line), compared to the algebraic (black dotted lines) and Gaussian bounds (brown dashed lines), respectively. The quantum bounds reach the algebraic upper bounds at $s=0$ for each case. In (a) and (b), the gap between quantum and classical bounds disappears at $s=-1$ and $s=-2$ for $\mathcal{J}_{s}$ and $\mathcal{J}_{s}^{\prime}$, respectively.} 
		\label{fig:QB1}
	\end{figure}

On the other hand, 
for a generalized distribution $s < 0$, we solve the eigenvalue problems in the basis of Fock states. To this aim, we first express $\hat{D} ( \alpha ) \hat{T} ( s ) \hat{D}^{\dag} ( \alpha )$ as
	\begin{equation}
		\hat{D} ( \alpha ) \hat{T} ( s ) \hat{D}^{\dag} ( \alpha ) = \sum_{n = 0}^{\infty} \sum_{m = 0}^{\infty} T_{n, m} \ketbra{n}{m},
	\end{equation}
where $T_{n,m} ( \alpha ) \equiv \mathrm{Tr} [ \ketbra{m}{n} \hat{D} ( \alpha ) \hat{T} ( s ) \hat{D}^{\dag} ( \alpha ) ]$ for $n \geq m$ is given by
	\begin{align} \label{eq:T}
		T_{n,m} ( \alpha ) & = \sqrt{\dfrac{m!}{n!}} \bigg( \dfrac{s+1}{s-1} \bigg)^{m} \exp \bigg( - \frac{2 | \alpha |^{2}}{1-s} \bigg) \nonumber \\
		& \times \bigg( \dfrac{2\alpha}{1-s} \bigg)^{n-m} L_{m}^{(n-m)} \bigg( \dfrac{4 | \alpha |^{2}}{1-s^{2}} \bigg),
	\end{align}
and $T_{n,m}$ for $n < m$ is obtained by $T_{m,n}^{*}$. Using Eq.~\eqref{eq:T}, we construct a density matrix for $\mathcal{H}_{s}$ and $\mathcal{H}_{s}^{\prime}$, and have obtained the maximum and minimum eigenvalues for each generalized quasiprobability function by extensive numerical optimizations. In Fig.~\ref{fig:QB1}, the maximal $\mathcal{J}_{s}$ and $\mathcal{J}_{s}^{\prime}$ for quantum states approach classical bounds as $s$ decreases, which implies that the whole set of quantum states detectable under our tests shrinks with $s$ decreasing.

For comparison, we also plot the Gaussian bounds obtained in the previous sections and the algebraic bounds obtainable from Eq.~\eqref{eq:SFAB}. The algebraically possible ranges of $\mathcal{J}_{s}$ and $\mathcal{J}_{s}^{\prime}$ are given by
	\begin{align}
		\frac{2s+4}{s-1} \leq \mathcal{J}_{s} \leq \frac{2s-4}{s-1} & \quad \mbox{for $-1 \leq s \leq 0$}, \nonumber \\
		-1 < \mathcal{J}_{s} < 3 & \quad \mbox{for $s < -1$},
	\end{align}
and
	\begin{align}
		\frac{s+3}{s-1} \leq \mathcal{J}_{s}^{\prime} \leq \frac{s-3}{s-1} & \quad \mbox{for $-1 \leq s \leq 0$}, \nonumber \\
		-1 < \mathcal{J}_{s}^{\prime} < 2 & \quad \mbox{for $s < -1$}.
	\end{align}
While the maximum and the minimum algebraic bounds are saturated by quantum states for $s=0$, there are gaps between algebraic and quantum bounds for a nonzero $s < 0$. As for the lower bounds of $\mathcal{J}_{s}$ and $\mathcal{J}_{s}^{\prime}$, the vanishing gap between algebraic and classical bounds clearly indicate that there is no quantum violation below $s=-1$. On the other hand, as for the upper bounds, we observe that the loss limits of $\mathcal{J}_{s}$ and $\mathcal{J}_{s}^{\prime}$ for the whole set of quantum states are identical to the loss limits for Gaussian states, 3 dB ($s=-1$) for $\mathcal{J}_{s}$ and 4.77 dB ($s=-2$) for $\mathcal{J}_{s}^{\prime}$, respectively. That is, below those limits, there exist some quantum states, both Gaussian (Sec. II C 2) and non-Gaussian, violating the classical bounds 2 and 1, respectively. 


	\begin{figure}[!t]
		\includegraphics[scale=0.45]{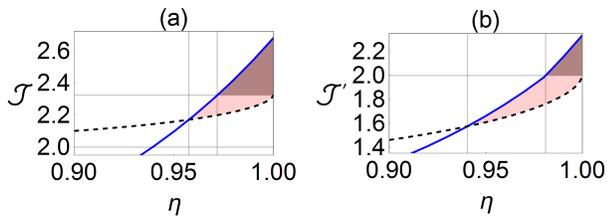}
		\caption{(Color online) (a, b) Optimal $\mathcal{J}_{s=1-\frac{1}{\eta}}$ and $\mathcal{J}_{s=1-\frac{1}{\eta}}^{\prime}$ for an even cat state $\ket{\psi_{\gamma}} = (2+2e^{-2|\gamma|^{2}})^{-1/2} ( \ket{\gamma} + \ket{-\gamma} )$ with $\gamma = 2$ (blue solid line), which are equivalent to $\mathcal{J}_{0}$ and $\mathcal{J}_{0}^{\prime}$ for a decohered state $\mathcal{L} [ \ketbra{\psi_{\gamma}}{\psi_{\gamma}} ]$ with loss parameter $1-\eta$ [Eq.~\eqref{eq:VRF}], respectively. On the other hand, black dashed lines denote the Gaussian bounds in Eq.~\eqref{eq:QNG1} [the same as Fig. 2(c) and 2(d)] for each $s = 1 - \frac{1}{\eta}$, above which genuine non-Gaussianity is detected. For a non-Gaussianity test, the detectable range of $\eta=\frac{1}{1-s}$ based on an $s$-parametrzied function for the original pure cat state (red shaded region) is larger than the range of $\eta$ based on the Wigner function for the decohered state under loss (brown shaded region). See main text.}
		\label{fig:CLD}
	\end{figure}

\section{Testing genuine non-Gaussianity}
As our tests are linear with respect to a convex mixture of quantum states, i.e., $\mathcal{J}_{s} [ \sum_{i} p_{i} \rho_{i} ] = \sum_{i} p_{i} \mathcal{J}_{s} [ \rho_{i} ]$, the gaps between the quantum and Gaussian bounds enable us to detect quantum non-Gaussianity. That is, if a given state is a mixture of Gaussian states, it must satisfy
	\begin{align} \label{eq:QNG1}
		-1 < & \mathcal{J}_{s} [ \rho_{\mathrm{MG}} = \sum_{i} p_{i} \sigma_{i} ] \leq \max_{\sigma} \mathcal{J}_{s}, \nonumber \\
		-1 < & \mathcal{J}_{s}^{\prime} [ \rho_{\mathrm{MG}} = \sum_{i} p_{i} \sigma_{i} ] \leq \max_{\sigma} \mathcal{J}_{s}^{\prime},
	\end{align}
where $\sigma_{i}$ is an arbitrary Gaussian state and the maximum values, $\max_{\sigma} \mathcal{J}_{s}$ and $\max_{\sigma} \mathcal{J}_{s}^{\prime}$, are shown in Figs. 2(c) and 2(d), respectively.
In particular, the case of $s=0$ gives
	\begin{align} \label{eq:QNGW}
		-1 < & \mathcal{J}_{0} [ \rho_{\mathrm{MG}} ] \leq \frac{8}{3^{9/8}}, \nonumber \\
		-1 < & \mathcal{J}_{0}^{\prime} [ \rho_{\mathrm{MG}} ] \leq 2.
	\end{align}

In Fig.~\ref{fig:CLD}, as an example, we plot the violation of inequalities \eqref{eq:QNG1} and \eqref{eq:QNGW} for an even cat state $\ket{\psi_{\gamma}} = (2+2e^{-2|\gamma|^{2}})^{-1/2} ( \ket{\gamma} + \ket{-\gamma} )$ under a lossy channel, which is represented by a Wigner function,
	\begin{align}
		& W_{\mathcal{L} [ \ketbra{\psi_{\gamma}}{\psi_{\gamma}} ]} ( q, p ; s=0 ) \nonumber \\
		& = \frac{2}{\pi} \frac{1}{1 + e^{-2\gamma^{2}}} e^{-2 q^{2} - 2 p^{2}} \{ e^{-2\eta \gamma^{2}} \cosh ( 4 \sqrt{\eta} \gamma q ) \nonumber \\
		& + e^{-2(1-\eta) \gamma^{2}} \cos ( 4 \sqrt{\eta} \gamma p ) \},
	\end{align}
with $\gamma$ real. We particularly compare (i) the Wigner-function-based tests $\mathcal{J}_{s=0}$ and $\mathcal{J}_{s=0}^{\prime}$ of the decohered state $\mathcal{L} [ \ketbra{\psi_{\gamma}}{\psi_{\gamma}} ]$ and (ii) the $s$-parametrized-function-based $\mathcal{J}_{s}$ and $\mathcal{J}_{s}^{\prime}$ of the original pure state $\ket{\psi_{\gamma}}$ with $s = 1 - \frac{1}{\eta}$, respectively. In Sec. II C 2, we have shown the equivalence of two nonclassicality tests---the Wigner-function test for the decohered state and the $s$-parametrized function test for the original state. In contrast, there exists inequivalence for non-Gaussianity tests---the latter detects non-Gaussianity in a broader parameter regime than the former. It is due to the fact that quantum non-Gaussianity bounds, $\max_{\sigma} \mathcal{J}_{s}$ and $\max_{\sigma} \mathcal{J}_{s}^{\prime}$, vary with the parameter $s$, whereas the nonclassicality bounds are the same regardless of $s$. On a practical side, it implies that the non-Gaussianity test for a decohered state under a lossy channel can be harder than expected from the analysis only based on the $s$-parametrized function of the original state.


	\begin{figure}[!t]
		\includegraphics[scale=0.45]{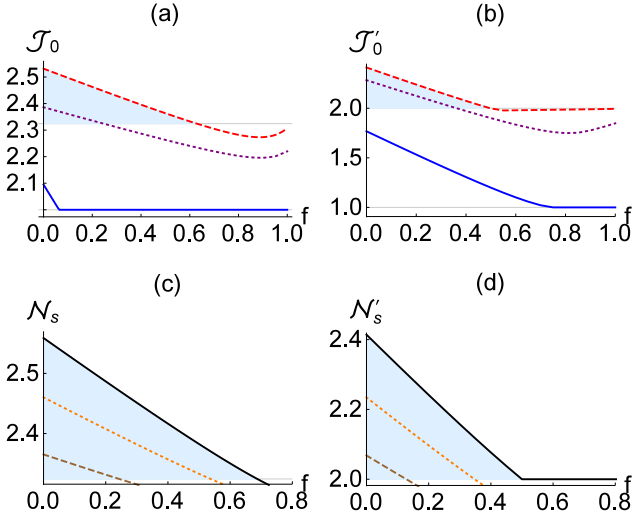}
		\caption{(Color online) (a, b) $\mathcal{J}_{0}$ and $\mathcal{J}_{0}^{\prime}$ optimized over the points of a rectangle (right triangle) for $\hat{S} \{ f \ketbra{0}{0} + (1-f) \ketbra{2}{2} \} \hat{S}^{\dag}$ as functions of the vacuum fraction $f$. Blue solid, purple dotted, and red dashed lines denote the case of squeezing applied to a given state, with squeezing $r = \{ 0, 0.5, 1 \}$ (a) and $r= \{ 0, 1, 2 \}$ (b), respectively. We find no violation of Eq.~\eqref{eq:QNGW} for an initial state ($r=0$), while a state with a sufficiently large squeezing violates Eq.~\eqref{eq:QNGW} (blue colored region) even when its Wigner function is positive ($f > \frac{1}{2}$). (c, d) $\mathcal{N}_{s}$ and $\mathcal{N}_{s}^{\prime}$ optimized over the points of a parallelogram (triangle) in Eqs. (41) and (42) for $f \ketbra{0}{0} + (1-f) \ketbra{2}{2}$ with respect to $f$. Black solid, orange dotted, brown dashed lines represent $s = \{ 0, -0.01, -0.02 \}$ (c) and $s = \{ 0, -0.02, -0.04 \}$ (d), respectively. The Wigner function $s=0$ shows the best performance to demonstrate quantum non-Gaussianity.}
		\label{fig:examples}
	\end{figure}

\subsection{Optimizing non-Gaussianity tests}
The test of genuine non-Gaussianity in Eq.~\eqref{eq:QNG1} can be further enhanced by applying squeezing $\hat{S}$ on a given state. Note that a mixture of Gaussian state remains to be a Gaussian mixture under Gaussian operations. Thus, if the state under squeezing shows violation of Gaussian bounds, the original state must be genuinely non-Gaussian. 
For example, we show the case of the mixed state $f\ketbra{0}{0} + (1-f) \ketbra{2}{2}$ in Figs.~\ref{fig:examples}(a) and~\ref{fig:examples}(b), which shows that the squeezing operation on the state enhances the detected region. 

\begin{figure}[!t]
		\includegraphics[scale=0.4]{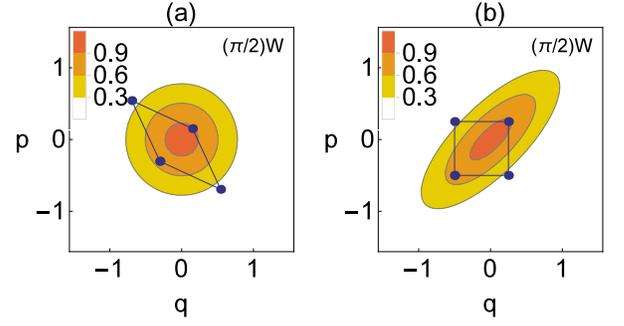}
		\caption{(Color online) A squeezing transformation converts (a) a parallelogram to (b) a rectangle while it squeezes the entire profile of Wigner function.} 
		\label{fig:squeezing}
	\end{figure}

For the case of a Wigner function, we may address the problem by using its interesting property \cite{Barnett}
	\begin{equation} \label{eq:121}
		W_{\hat{S} \rho \hat{S}^{\dag}} ( \alpha; 0 ) = W_{\rho} ( S [ \alpha ]; 0 ),
	\end{equation}
where $S [ \alpha ] = \alpha \cosh r + \alpha^{*} e^{2i\phi} \sinh r$ is the transformation of phase-space points due to the squeezing operation $\hat{S}$. As we illustrate in Fig.~\ref{fig:squeezing}, under the squeezing transformation $S [ \alpha ]$, a parallelogram in an unsqueezed profile corresponds to a rectangle in a squeezed profile. 
Importantly, it means that we do not need to implement a squeezing operation on a given state in order to have an enhanced test. Instead, we may simply choose the four points at the vertices of the parallelogram corresponding to the squeezing operation and do our tests for the given initial state. 

On the other hand, for a nonzero $s<0$, the above argument is not directly applicable. This is because the $s$-parametrized function of a state after squeezing is not simply obtained by reshaping of the original $s$-parametrized function (squeezing of the profile). Instead, we find the following identity (with its proof in the Appendix):
\begin{equation} \label{eq:SGR}
		\frac{\pi (1-s)}{2} W_{\rho} ( S [ \alpha ]; s ) = \frac{\pi}{2} W_{\mathcal{L}^{\prime} [ \hat{S} \rho \hat{S}^{\dag} ]} \bigg( \frac{\alpha}{\sqrt{1-s}}; 0 \bigg),
	\end{equation}
where $\mathcal{L}^{\prime} [ \rho ]$ represents a beam-splitter interaction with transmittance $\eta = \frac{1}{1-s}$ between a quantum state $\rho$ and a squeezed vacuum $\hat{S} \ketbra{0}{0} \hat{S}^{\dag}$. 
Therefore, all $s$-parametrized functions taken at the vertices of a paralleogram can be understood as the Wigner function taken at the vertices of a rectangle for the decohered state $\mathcal{L}^{\prime} [ \hat{S} \rho \hat{S}^{\dag} ]$. As the whole process $\mathcal{L}^{\prime} [ \hat{S} \rho \hat{S}^{\dag} ]$ is Gaussian, it does not create non-Gaussianity. We then use the Gaussian bounds obtained for $s=0$ in order to detect genuine non-Gaussianity under the $s$-parametrized functions. 

We thus propose enhanced non-Gaussianity tests for an arbitrary $s$ as
	\begin{align}
		-1 < & \mathcal{N}_{s} [ \rho_{\mathrm{MG}} ] \leq \frac{8}{3^{9/8}}, \nonumber \\
		-1 < & \mathcal{N}_{s}^{\prime} [ \rho_{\mathrm{MG}} ] \leq 2,
	\end{align}
with
	\begin{align} \label{eq:NP}
		\mathcal{N}_{s} [ \rho ] & = \frac{\pi (1-s)}{2} \{ W_{\rho} ( S [ q_{0} + i p_{0} ]; s ) + W_{\rho} ( S [ q_{1} + i p_{0} ]; s ) \nonumber \\
		& + W_{\rho} ( S [ q_{0} + i p_{1} ]; s ) - W_{\rho} ( S [ q_{1} + i p_{1} ]; s ) \},   
	\end{align}
and
	\begin{align} \label{eq:NP1}
		\mathcal{N}_{s}^{\prime} [ \rho ] & = \frac{\pi (1-s)}{2} \{ W_{\rho} ( S [ q_{1} + i p_{0} ]; s ) + W_{\rho} ( S [ q_{0} + i p_{1} ]; s ) \nonumber \\
		& - W_{\rho} ( S [ q_{1} + i p_{1} ]; s ) \},
	\end{align}
which take into account the vertices of parallelogram and triangle, respectively, with $S [ \alpha ] = \alpha \cosh r + \alpha^{*} e^{2i\phi} \sinh r$. 
Note that any values of $r$ and $\phi$ for the squeezing transformation $S [ \alpha ]$ can be used in Eqs.~\eqref{eq:NP} and \eqref{eq:NP1}, as squeezing does not create non-Gaussianity. As an illustration, we show the case of the state $f\ketbra{0}{0} + (1-f) \ketbra{2}{2}$ in Figs.~\ref{fig:examples}(c) and~\ref{fig:examples}(d) using the above tests. 


\subsection{Testing noisy non-Gaussian states}

We here investigate our test of genuine non-Gaussianity for noisy non-Gaussian states. In particular, when a single-mode state undergoes a lossy Gaussian channel, the corresponding Wigner function becomes positive definite if its transmission rate $\eta$ is below 50\%. It is then interesting to know whether our non-Gaussianity test can detect noisy non-Gaussian states for $\eta<0.5$. For this purpose, we consider a class of superposition states $|\Psi_N\rangle=\sum_{n=0}^NC_n|n\rangle$ ($N$: truncation number) \cite{LN} as an input to a lossy channel. We then test the genuine non-Gaussianity for the output decohered states using the Wigner function ($s=0$) or the $Q$ function ($s=-1$) in Eqs.~\eqref{eq:NP} and~\eqref{eq:NP1}. 
	
	\begin{figure}
		\includegraphics[scale=0.55]{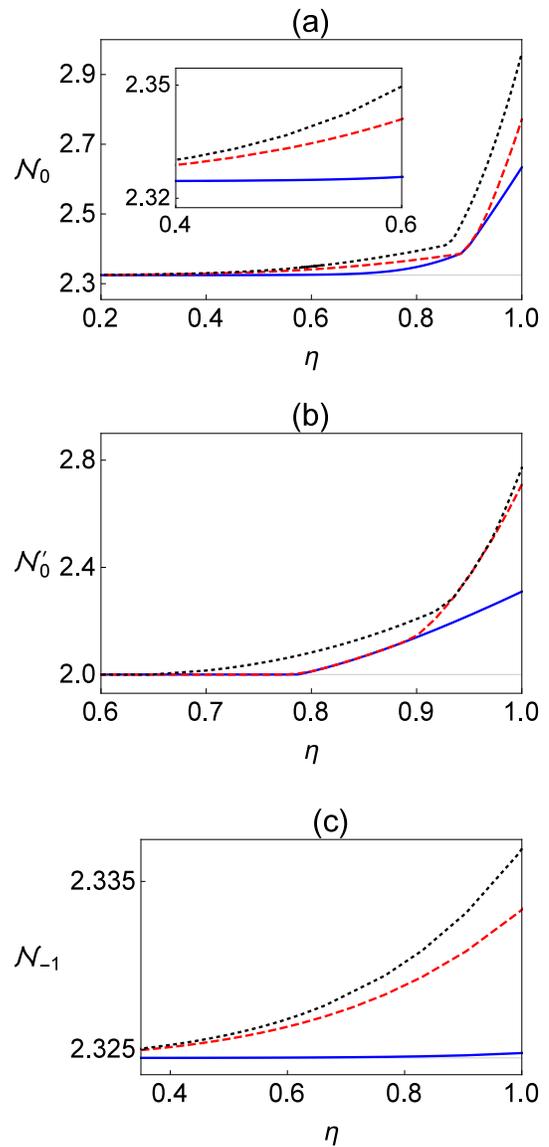}
		\caption{(Color online) (a) $\mathcal{N}_{s=0}$ (parallelogram test), (b) $\mathcal{N}_{s=0}^{\prime}$ (triangle test), and (c) $\mathcal{N}_{s=-1}$ (parallelogram test), respectively, optimized over a superposition state $\sum_{n=0}^NC_n|n\rangle$ for each $\eta$ (transmission rate) under a loss channel. The curves from top to bottom represent the case of truncation number $N=3$ (black dotted), $N=2$ (red dashed) and $N=1$ (blue solid). $\mathcal{N}_{s}>\frac{8}{3^{9/8}}\approx2.324$ and $\mathcal{N}_{s}^{\prime}>2$, respectively, represent genuine non-Gaussianity of the noisy non-Gaussian states.}
		\label{fig:OT}
	\end{figure}	
	
In Fig.~\ref{fig:OT} (a), we plot $\mathcal{N}_{s=0}$ (parallelogram test) for the superposition states $\sum_{n=0}^NC_n|n\rangle$ with varying truncation number $N=1$ (blue solid), $N=2$ (red dashed), and $N=3$ (black dotted). 
For each $\eta$, we optimize the coefficients $C_n$ to show a maximal $\mathcal{N}_{0}$ in the figures. We see that $\mathcal{N}_{0}$ is above the Gaussian bound	$\frac{8}{3^{9/8}}$ even below $\eta=0.5$ (magnified view in the inset), demonstrating a successful detection of non-Gaussianity for positive Wigner functions. The plots indicate that there exist some superposition states for each $N$ whose non-Gaussianity can be detected under transmission below 50\%. We have numerically found that $\mathcal{N}_{0}>\frac{8}{3^{9/8}}$ appears at a very low $\eta$. 
If we take into account the violation of size $\sim0.001$, the critical $\eta$ turns out to be $\eta_c=0.590$ ($N=1$), $\eta_c=0.238$ ($N=2$), and $\eta_c=0.221$ ($N=3$), respectively.

On the other hand, under the triangle test $\mathcal{N}_{0}^{\prime}$, the detection becomes less effective as shown in Fig.~\ref{fig:OT} (b). In particular, the critical $\eta$ does not go below 0.5 in contrast to the rectangle test $\mathcal{N}_{0}$ in Fig.~\ref{fig:OT} (a). Moving on to the $Q$-function tests ($s=-1$), they are naturally less powerful than the Wigner-function test ($s=0$). However, they can also manifest genuine non-Gaussianity below $\eta=0.5$ under the paralleogram test as shown in Fig.~\ref{fig:OT} (c), with critical values $\eta_c=0.476$ ($N=2$) and $\eta_c=0.442$ ($N=3$) at the violation of size 0.001 (no violation for $N=1$). We do not show the case of triangle test $\mathcal{N}_{s=-1}^{\prime}$, which does not detect non-Gaussianity. Thus, for the class of noisy non-Gaussian states considered here, we find that the parallelogram test with four phase-space points is more powerful than the triangle test with three phase-space points.

\section{Connection between single-mode and two-mode tests}
Finally, we address the connection between our single-mode nonclassicality test and the BW two-mode nonlocality test more directly. 
We start our discussion with a single-mode state $\rho$ that satisfies a condition
	\begin{equation} \label{eq:CST}
		\mathcal{N} [ \rho ] > 2 e^{\frac{1}{2} \max [ | D_{a} |, | D_{b} |  ]^{2}},
	\end{equation} 
where $\mathcal{N} [ \rho ]$ is the quantity defined in Eq.~\eqref{eq:NP} ($s=0$), and $D_{a} = S [ q_{1} + i p_{1} ] - S [ q_{0} + i p_{0} ]$ and $D_{b} = S [ q_{1} + i p_{0} ] - S [ q_{0} + i p_{1} ]$ correspond to the diagonal lengths of a parallelogram with the transformation $S [ \alpha ] = \alpha \cosh r + \alpha^{*} e^{2i\phi} \sinh r$. 
For the case of taking $r=0$, the four points simply form a rectangle; then $\mathcal{N} [ \rho ]$ has a classical bound 2. In other words, Eq.~\eqref{eq:CST} becomes a nonclassicality test and the quantity $2e^{\frac{1}{2} \max [ | D_{a} |, | D_{b} |  ]^{2}}>2$ represents a degree of nonclassicality. On the other hand, for the case of taking $r>0$, Eq.~\eqref{eq:CST} may not be immediately regarded as a nonclassicality test, since a squeezing operation can create nonclassicality to the input state. Nevertheless, we show below that if a state $\rho$ satisfies the condition in Eq.~\eqref{eq:CST} for any values of $r$ and $\phi$, it manifests nonlocality under the BW test  
	\begin{align}
		\mathcal{B} [ \rho_{AB} ] & \equiv \frac{\pi^{2}}{4} \{ W_{\rho_{AB}} ( \alpha_{0}, \beta_{0} ) + W_{\rho_{AB}} ( \alpha_{1}, \beta_{0} )  \nonumber \\
		& + W_{\rho_{AB}} ( \alpha_{0}, \beta_{1} ) - W_{\rho_{AB}} ( \alpha_{1}, \beta_{1} ) \}>2,
	\end{align}
by mixing it with a vacuum at a 50:50 beam splitter.

	\begin{figure}[!t]
		\includegraphics[scale=0.5]{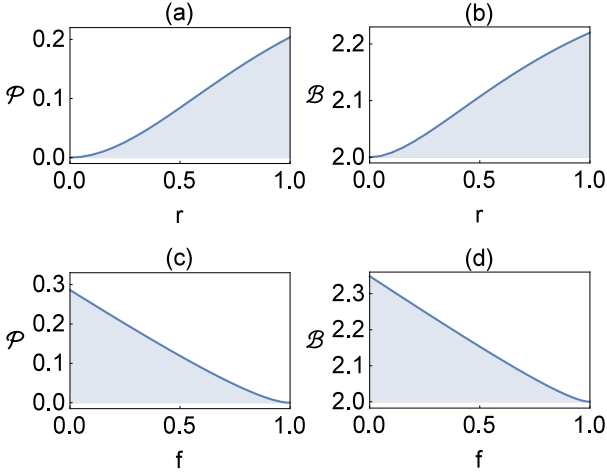}
		\caption{(Color online) (a, b) Optimal $\mathcal{P}\equiv \mathcal{N} [ \rho ]- 2 e^{\frac{1}{2} \max [ | D_{a} |, | D_{b} | ]^{2}}$ and $\mathcal{B}$ for squeezed states as functions of squeezing strength $r$. (c, d) Optimal $\mathcal{P}$ and $\mathcal{B}$ for states $f \ketbra{0}{0} + (1-f) \ketbra{2}{2}$ as functions of the vacuum fraction $f$. Every $\mathcal{P}>0$ in our single-mode test is monotonically connected to the violation of the nonlocality test $\mathcal{B} > 2$ in phase space.}
		\label{fig:CST}
	\end{figure}

It has been known that a single-mode nonclassical state can be turned into an entangled state using a beam-splitter setting \cite{Kim2002,Wolf2003,Asboth2005,Nha2008,Tahira2009}.
By mixing a single-mode state $\rho$ with a vacuum at a 50:50 beam splitter, we obtain a two-mode Wigner function as
	\begin{equation}
		W_{\rho^{\prime}_{AB}} ( \alpha, \beta ) = W_{\rho} \bigg( \frac{\alpha + \beta}{\sqrt{2}} \bigg) W_{\ketbra{0}{0}} \bigg( \frac{- \alpha + \beta}{\sqrt{2}} \bigg),
	\end{equation}
where $W_{\ketbra{0}{0}}(\gamma)\equiv\frac{2}{\pi}e^{-2|\gamma|^2}$ is the Wigner function of a vacuum state. 
Let us assume that Eq.~\eqref{eq:CST} is satisfied for the state $\rho$. 
Using a short-hand notation $S [ q_{i} + i p_{j} ]\equiv S_{ij}$, we choose the phase-space points for the BW test as
	\begin{align}
		\alpha_{0} & = \frac{1}{2 \sqrt{2}} ( 2 S_{00} - S_{10} + S_{01} ), \nonumber \\
		\alpha_{1} & = \frac{1}{2 \sqrt{2}} ( 2 S_{11} + S_{10} - S_{01} ), \nonumber \\
		\beta_{0} & = \frac{1}{2 \sqrt{2}} ( 2 S_{00} + S_{10} - S_{01} ), \nonumber \\
		\beta_{1} & = \frac{1}{2 \sqrt{2}} ( 2 S_{11} - S_{10} + S_{01} ),
	\end{align}
and obtain
	\begin{align}
		\mathcal{B} [ \rho^{\prime}_{AB} ] & = \frac{\pi}{2} \{ W_{\rho} ( S_{00} ; 0 ) - W_{\rho} ( S_{11} ; 0 ) \} e^{- \frac{1}{2} | D_{b} |^{2}} \nonumber \\
		& + \frac{\pi}{2} \{ W_{\rho} ( S_{10} ; 0 ) + W_{\rho} ( S_{01} ; 0 ) \} e^{- \frac{1}{2} | D_{a} |^{2}} \nonumber \\
		& \geq \mathcal{N} [ \rho ]  e^{- \frac{1}{2} \max [ | D_{a} |, | D_{b} | ]^{2}} \nonumber \\
		& > 2,
	\end{align}
with the identity $S_{00}+S_{11}=S_{10}+S_{01}$. 	
We have above used that $W_{\rho} ( S_{00} ; 0 ) - W_{\rho} ( S_{11} ; 0 )$ and $W_{\rho} ( S_{10} ; 0 ) + W_{\rho} ( S_{01} ; 0 )$ are both positive if $\mathcal{N} [ \rho ] > 2$; if the former is negative, then $\mathcal{N} [ \rho ] < 2$ due to the constraint $\frac{\pi}{2} \{ W_{\rho} ( S_{10} ; 0 ) + W_{\rho} ( S_{01} ; 0 ) \} \leq 2$.

As a by-product, we now confirm that Eq.~\eqref{eq:CST} represents a nonclassicality test for any values of $r$ and $\phi$. This is because nonclassicality is a necessary condition to make a nonlocal resource in the beam-splitter setting. 
As an illustration, in Fig.~\ref{fig:CST} we plot the optimal value of $\mathcal{P} \equiv \mathcal{N} [ \rho ] - 2 e^{\frac{1}{2} \max [ | D_{a} |, | D_{b} | ]^{2}}$ and the corresponding BW quantity $\mathcal{B}$ for squeezed states and noisy Fock states, respectively, which clearly manifests that our single-mode test can represent a nonlocal resource in phase space.

\section{Conclusions and Remarks}
We have proposed Bell-type tests of non-classicality and non-Gaussianity using generalized phase-space distributions. Our rectangle and right-triangle tests are capable of detecting a wide range of nonclassical states including mixed Gaussian squeezed states and non-Gaussian states that possess even positive-definite distributions in phase space. 
For nonclassicality tests, we have identified the ultimate limits of 3-dB (transmittance $\eta=\frac{1}{2}$) and 4.77-dB ($\eta=\frac{1}{3}$) under a loss channel with our approach employing four and three phase-space points, respectively. 
Furthermore, we have shown that our test can be robust against experimental imperfections, including finite data acquisition and nonoptimal choice of phase-space points. 

As our nonclassicality tests set bounds for all Gaussian states and their mixtures, they can be further used as criteria to detect genuine quantum non-Gaussianity, which is known to be a crucial resource for numerous quantum tasks. We have obtained the Gaussian bounds for all tests using the $s$-parametrized distributions. Remarkably, we have shown how the rectangle and the right-triangle tests can be generalized to the parallelogram and the triangle tests, respectively, which is essentially equivalent to a squeezing operation on a given state, thereby enhancing the successful detection of genuine non-Gaussianity. 
Importantly from a practical point of view, the optimized parallelogram and triangle tests do not require an actual realization of squeezing operation, as the choice of appropriate phase-space points serves the purpose. 

Employing four and three phase-space point tests have their own advantages in our tests. For instance, we have found that three-point tests are more advantageous in detecting the nonclassicality of mixed Gaussian states. On the other hand, four-point tests are more powerful in detecting genuine non-Gaussianity of non-Gaussian states under a loss channel even below $\eta=\frac{1}{2}$. 


Detecting nonclassicality and quantum non-Gaussianity using positive valued points in phase space requires a certain constraint, e.g., predetermined position (origin) with energy constraint \cite{Genoni2013} and a designated shape (tetragon or triangle) in our tests \cite{Park2015}. It may be possible to enhance our tests by exploiting more constraints such as the input-state energy, the length of vertices, the area of a polygon, etc. In this respect, we have addressed an example that the diagonal lengths in the parallelogram test can be used as valuable information to connect our single-mode test and the BW nonlocality test in phase space. As nonlocality tests have evolved from simple to more complex tests \cite{Brunner2014}, we hope our work can further stimulate efforts for more elaborate tests to detect nonclassical states and find useful resources for genuine multi-mode nonlocality tests.

\section*{Acknowledgment}
We acknowledge support through NPRP, Grant No. 7-210-1-032, from the Qatar National Research Fund.

\bibliographystyle{apsrev}

\section*{Appendix A. Proof of Eq.~\eqref{eq:SGR}}
\begin{widetext}
Applying a linear transformation in Eq.~\eqref{eq:GC}, we first have
	\begin{equation} \tag{A1}
		W_{\rho} ( \alpha \cosh r + \alpha^{*} e^{2i \phi} \sinh r; s ) = \frac{2}{\pi ( - s )} \int d^{2} \beta W_{\rho} ( \beta; 0 ) \exp \bigg( - \frac{2 | \alpha \cosh r + \alpha^{*} e^{2i \phi} \sinh r - \beta |^{2}}{(-s)} \bigg).
	\end{equation}
Changing the variable $\beta$ to $\beta^{\prime} \cosh r + \beta^{\prime} e^{2i \phi} \sinh r$ and using the property $W_{\rho} ( \alpha \cosh r + \alpha^{*} e^{2i \phi} \sinh r; 0 ) = W_{\hat{S} \rho \hat{S}^{\dag}} ( \alpha; 0 )$, we obtain
	\begin{align} \label{eq:AA1}
		& W_{\rho} ( \alpha \cosh r + \alpha^{*} e^{2i \phi} \sinh r; s ) \nonumber \\
		& = \frac{2}{\pi ( - s )} \int d^{2} \beta^{\prime} W_{\rho} ( \beta^{\prime} \cosh r + \beta^{\prime *} e^{2i \phi} \sinh r; 0 ) \exp \bigg( - \frac{2 | ( \alpha - \beta ) \cosh r + ( \alpha - \beta )^{*} e^{2i \phi} \sinh r |^{2}}{(-s)} \bigg) \nonumber \\
		& =  \frac{2}{\pi ( - s )} \int d^{2} \beta^{\prime} W_{\hat{S} \rho \hat{S}^{\dag}} ( \beta^{\prime}, 0 ) \exp \bigg( - \frac{2 | ( \alpha - \beta ) \cosh r + ( \alpha - \beta )^{*} e^{2i \phi} \sinh r |^{2}}{(-s)} \bigg). \tag{A2}
	\end{align}
As the interaction between a quantum state $\rho$ and a reservoir $\sigma$ is expressed by a beam-splitter with transmittance $\eta$, we have
	\begin{align} \label{eq:AA2}
		W_{\mathcal{L}^{\prime} [ \rho ]} ( \alpha; 0 ) & = \int d^{2} \beta W_{\rho} ( \sqrt{\eta} \alpha - \sqrt{1-\eta} \beta; 0 ) W_{\sigma} ( \sqrt{1-\eta} \alpha + \sqrt{\eta} \beta; 0 ) \nonumber \\
		& = \frac{1}{1-\eta} \int d^{2} \beta^{\prime} W_{\rho} ( \beta^{\prime}; 0 ) W_{\sigma} \bigg[ \sqrt{\frac{\eta}{1-\eta}} \bigg( \frac{\alpha}{\sqrt{\eta}} - \beta^{\prime} \bigg); 0 \bigg]. \tag{A3}
	\end{align}
Comparing Eqs.~\eqref{eq:AA1} and \eqref{eq:AA2}, we see that the reshaping (squeezing) of a profile in phase space is related to a beam-splitter interaction between a squeezed quantum state $\hat{S} \rho \hat{S}^{\dag}$ and a squeezed vacuum reservoir $\hat{S} \ketbra{0}{0} \hat{S}^{\dag}$ with transmittance $\eta = \frac{1}{1-s}$,
	\begin{equation} \tag{A4}
		W_{\rho} ( \alpha \cosh r + \alpha^{*} e^{2i \phi} \sinh r; s ) = \frac{1}{1-s} W_{\mathcal{L}^{\prime} [ \hat{S} \rho \hat{S}^{\dag} ]} \bigg( \frac{\alpha}{\sqrt{1-s}}; 0 \bigg).
	\end{equation}
\end{widetext}
\end{document}